\documentclass[aps,twocolumn, pra,showpacs,floatfix]{revtex4-1}
\usepackage{times}
\usepackage{amsmath}
\usepackage{amsfonts}
\usepackage{amssymb}
\usepackage{graphicx}
\usepackage{natbib}
\usepackage{color}
\begin{document}
\title{Quantum interference controlled resonance profiles: \\From lasing without inversion to photo-detection}
\author{Konstantin E. Dorfman$^{1}$}
\email{Corresponding author. Email: dorfman@physics.tamu.edu}
\author{Pankaj K. Jha$^{1}$}
\email{Email: pkjha@physics.tamu.edu}
\author{Sumanta Das$^{2}$}
\email {Email: sumanta.das@mpi-hd.mpg.de}
\affiliation{$^{1}$Institute for Quantum Science and Engineering and Department of Physics and 
Astronomy, Texas A$\&$M University, College Station, TX 77843-4242, USA\\
$^{2}$Max-Planck-Institut f\"{u}r Kernphysik, Saupfercheckweg 1, 69117 Heidelberg, Germany} 
\date{\today}
\begin{abstract}
In this work we report a quantum interference mediated control of the resonance profiles in a generic three-level system and investigate its effect on key quantum interference (QI) phenomena. Namely in a three level configuration with doublets in the ground or excited states, we show control over enhancement and suppression of the  emission (absorption) profiles. This is achieved by manipulation of the strength of QI and the energy spacing of the doublets.  We analyze the application of such QI induced control of the resonance profile in the framework of two limiting cases of lasing without inversion and photo-detection.
\end{abstract}
\maketitle
\section{Introduction}
Study of quantum interference (QI) had led to the discovery of numerous fascinating phenomena in various  type of systems ranging from atoms to biomolecules \cite{Scub, Ficb, Fla09}.  In atomic systems for example, one of the earliest known effect of QI is the modification of the absorption profiles that comes about due to interference among the bound-bound and bound-continuum transitions, a phenomenon now called Fano interference \cite{Fan61}. Agarwal \cite{Gsab} later showed how QI among decay pathways can lead to generation of coherence and population trapping in a multi-level atomic configuration. A counter-intuitive application of such \textit{Agarwal-Fano QI} was discovered by Harris in the form of inversion-less lasing (LWI) \cite{Har89}. This non-energy conserving phenomena had thereof lead to several theoretical investigations \cite{Scu89, Koc92, Man92} and experimental demonstration~\cite{LWI12, LWI4,LWI5}. Furthermore, during the past decade study of QI effects has been extended to tailored semiconductor nanostructures like quantum wells and dots due to coherent resonant tunneling owing to their potential applications in photo-detection \cite{Woj08, Vas11}, lasing \cite{Cap97,Sch97}, quantum computing and quantum circuitry \cite{And10, Gib11}. 

In the seminal work of Scully \cite{Scu10} it was shown that coherence induced by external source can break the detailed balance between emission and absorption and  enhance, in principle, the quantum efficiency of a photovoltaic cell. Ref \cite{Scu10} demonstrated the role of quantum coherence in a simple way. In a recent work we showed that coherence induced by QI can enhance the power of the Photocell and Laser Quantum Heat Engines \cite{Dor11, Dor111} following the earlier work on Photo-Carnot Engine enhanced by quantum coherence \cite{Scu03}. The main idea is that the quantum coherence induced by either an external drive or QI among the decay paths alters the detailed balance between emission and absorption and can enhance the efficiency of the system compared to that without quantum coherence. In the case of photovoltaic cells quantum coherence leads to suppression of radiative recombination  \cite{Scu10}  or enhancement of absorption \cite{Dor111} and thus, increase of the efficiency. Furthermore, the results of the Ref. \cite{Scu10} have initiated debates about the principle issues.  In his article \cite{Kirk11} Kirk attempts to investigate the limits of Ref. \cite{Scu10} and, in particular, argues that Fano interference does not break detailed balance of the photocell. Note, that noise induced coherence via Fano interference was later shown to indeed enhance the balance breaking in photovoltaics where it leads to increase in power \cite{Scu11, Chap11, Dor11, Dor111}.

These investigations have hence generated renewed interest in the fundamental question of noise 
induced interference effects on the emission and absorption profile of an atom or atom like system 
(excitons in quantum wells or dots) \cite{Wan11}. As such, we in this paper undertake a thorough 
theoretical investigation of the vacuum induced interference effects on the resonance line profiles 
of a three level system with doublets in ground (excited) state configuration (see Fig. 1). Our analysis is quite general and applies to atoms, molecules as well as quantum wells and dots. We study the time profile of absorption and emission probabilities and derive its close form expression in the steady state regime. In the present work we use a simple probability amplitude method to calculate the resonance profiles since the states involved in calculation have zero photon occupation number. The latter is equivalent to the density matrix formalism usually used in this type of problems \cite{Dor11, Dor111}.

The probabilities of emission and absorption are found 
to have strong functional dependence on the the energy spacing between the doublets ($2\Delta$) and interaction strength $p$. In the case of atomic system $p$ is governed by a mutual orientation of dipole moments. In semiconductor systems $p$ has a meaning of the phase shift acquired by the wave function between two interfering pathways. This thus provides us with two different parameter by which we can regulate 
the QI in the system. For example, we show that depending on the choice of energy spacing between the doublets 
compared to spontaneous decay rate we can use destructive interference to achieve either LWI by 
enhancing the emission or photodetectors and interferometers by reducing emission and enhancing 
absorption. Moreover depending on the $p$ we can manipulate the 
interference type from destructive to constructive which can significantly alter the resonance 
profiles (see Fig. 3.)

The outline of the paper is as follows. In section II, we present our theoretical model of 
three-level system with the doublet in the ground state and calculate the expression for 
the probability of emission $P_{emiss}$ and absorption $P_{abs}$ in the long time limit 
$t \gg \gamma^{-1}, \Gamma^{-1}$. 
Furthermore,  we also give results for emission and absorption probabilities for a three-level 
configuration with upper state doublet. In section III, the functional form of the ratio $P_{emiss}/P_{abs}$ is presented to quantify its dependence on the dipole alignment 
parameter $p$, energy spacing $(2\Delta)$ and the radiative decay rates $\Gamma,\gamma$. We discuss our results and propose potential application of our model to enhancement of emission in LWI configuration, enhancement of absorption for photodetectors, measurement of high to moderate magnetic field intensities and observation of QED results on quantum beats in the semi-classical regime. Finally in section IV we conclude by summarizing our findings.

\section{Theoretical model} 
In order to investigate the effect of QI on the emission and absorption profile of an atomic, molecular 
or semiconductor system we consider a  three level configuration with a ground state doublet $|v_{1,2}\rangle$ 
and excited state $|c\rangle$ (see Fig. 1a). The three level system is excited by coherent field with 
the central frequency $\nu$ so that the energies of state  $|v_{1,2}\rangle$ are related to $|c\rangle$ 
as $\nu \pm \Delta$, where $\Delta$ half of the energy spacing between the ground state doublet. The ground state doublet $|v_{1,2}\rangle$ 
decays to the reservoir state $R_v$ with the rate  $2\gamma_{1,2}$ respectively and the excited state 
decays to the reservoir state $R_c$ with the decay rate $2\Gamma$. Furthermore, states $|v_{1,2}\rangle$ can 
represent either Zeeman sub-levels in atoms, vibrational levels within electronic band in molecules or 
intrasubband in semiconductor. Since the typical relaxation rate of electronic (intersubband) transition 
is much smaller than that of vibrational (intrasubband), we neglect direct decay process between level 
$|c\rangle$ and $|v_{1,2}\rangle$. Note that  the decay of ground state doublets $|v_{1,2}\rangle$ to 
the same state $|R_v\rangle$ leads to a vacuum induced coherence among them. The physics of this coherence 
is attributed to the \textit{Agarwal-Fano QI} of the transition amplitudes among the decay pathways. 
Note that analysis presented below is valid for the system with excited state doublet $|c_{1,2}\rangle$ 
and single ground state $|v\rangle$ (as per Fig.1b, see discussion). We will show later  that such QI plays a major role in the line profiles of an atomic system~\cite{Har89, Ima95}.  
The time dependent amplitudes of the states $|v_{1,2}\rangle$ and $|c\rangle$ essentially exhibits the 
effect of coherence 
on the dynamics of the system. The probability amplitude method can be applied in the present system since states $|v_{1,2}\rangle$ and $|c\rangle$ have zero photon occupation number. Solving the time dependent Schr\"odinger equation, the 
dynamical evolution of the probability amplitudes  $v_{1,2}$ and $c$ of finding system in corresponding states $|v_{1,2}\rangle$ and $|c\rangle$ (i.e. states with zero photons) in Weisskopf-Wigner approximation is given by
\begin{figure}[t]\label{fig:3levvv}
\centerline{\includegraphics[height=6.1cm,width=0.49\textwidth,angle=0]{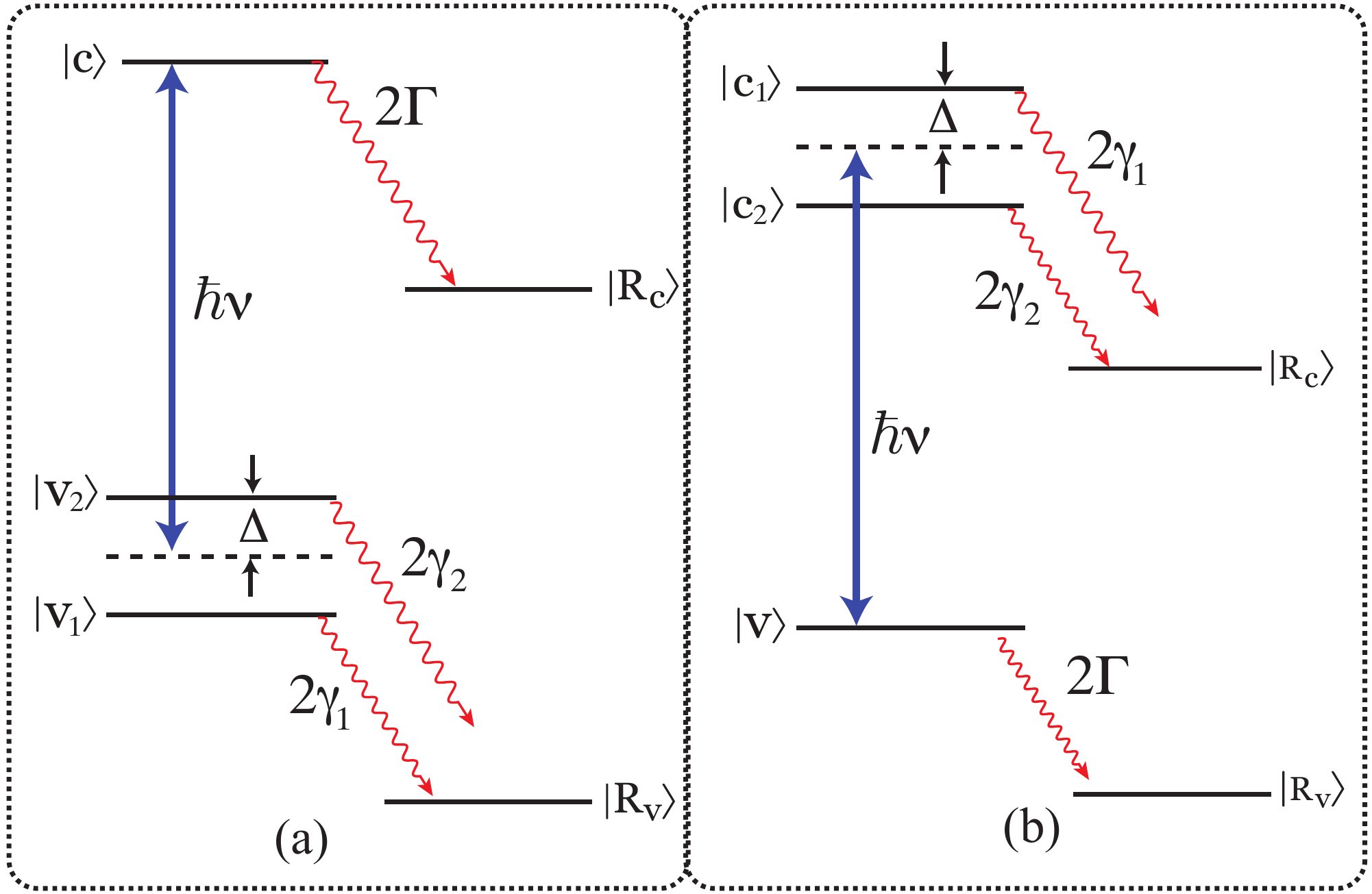}}
\caption{(Color online) The scheme of the three level system with the doublet in the ground state (a) and in the excited state (b).
Radiative decay from the doublet states to the reservoir is $2\gamma$ while the excited (ground) state to the reservoir
is $2\Gamma$.}
\end{figure}
\begin{equation}
\dot{v}_2(t)=-(\gamma_2+i\Delta)v_2(t)-p\sqrt{\gamma_1\gamma_2} v_1(t)-i\Omega_2c(t) \label{eq:a2},
\end{equation}
\begin{equation}
\dot{v}_1(t)=-(\gamma_1-i\Delta)v_1(t)-p\sqrt{\gamma_1\gamma_2} v_2(t)-i\Omega_1c(t) \label{eq:a1},
\end{equation}
\begin{equation}
\dot{c}(t)=-i\Omega_2v_2(t)-i\Omega_1v_1(t)-\Gamma c(t) \label{eq:b},
\end{equation}
where $\Omega_{1,2}=\wp_{1,2} \mathcal{E}_0/2\hbar$ and $\wp_{1,2}$ are the respective Rabi frequencies 
and dipole moments of the corresponding transitions $|v_{1,2}\rangle \leftrightarrow |c\rangle$ with $\mathcal{E}_0$ 
being the amplitude of the applied electric field. The term $p\sqrt{\gamma_{1}\gamma_{2}}$ arises due to QI of the decay pathways of the ground state doublet. It is clearly seen from the above set of 
equations that this term for $p \neq 0$ couples the amplitudes of the states $v_1$ and $v_2$. Such a 
coupling is known as \textit{Agarwal-Fano coupling} in the literature \cite{ScuL07} and have several 
implications ranging from superradiance \cite{Scu06, Das08} and entanglement \cite{Das08} to quantum solar cells 
\cite{Scu10, Dor11, Dor111}. The interference strength is typically determined in terms of the 
relative orientation of the dipole moments of the decay transitions and is given by coefficient a $p$ as,
\begin{equation}
p=\frac{\vec{\wp}_{v_{1}R_{v}}\cdot\vec{\wp}_{v_{2}R_{v}}}{|\vec{\wp}_{v_{1}R_{v}}||\vec{\wp}_{v_{2}R_{v}}|}
\end{equation}
\begin{figure*}[!ht]\label{fig:3levvv}
\centerline{\includegraphics[height=7.2cm,width=0.99\textwidth,angle=0]{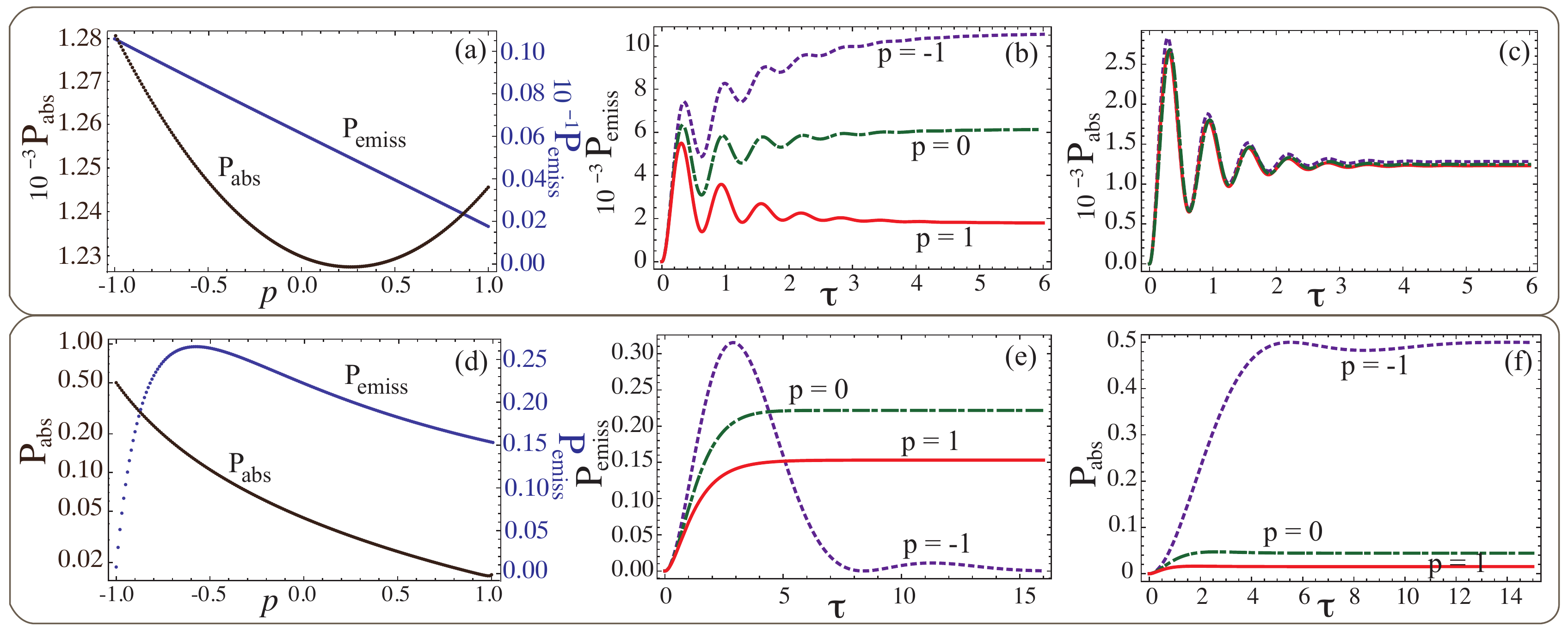}}
\caption{(Color online) Steady state (a,d) and temporal evolution (b,c,e,f) of the emission and absorption probability for 
the three level model with the doublet in the ground state. (a,d) shows the effect of the parameter 'p' on 
the steady state values of the probability of emission and absorption. (b,e) shows the temporal behavior of 
the probaility of emisison for three choices of 'p'. (c,f) shows the temporal behavior of the probability of 
absorption for the same choices of p as in (b,e). For numerical simulation we took, 
$\gamma=1$, $\Omega_{1}=\Omega_{2}=0.3\gamma$, $\Gamma=0.4\gamma$ and $\Delta =10\gamma$ for (a,b,c) 
and $\Delta =0.01\gamma$ for (d,e,f).}
\end{figure*}
\noindent where $\vec{\wp}_{v_{1}R_{v}}$ and $\vec{\wp}_{v_{2}R_{v}}$ are the dipole moment corresponding 
to the transition $|v_{1} \rangle \leftrightarrow |R_{v}\rangle$ and $|v_{2} \rangle \leftrightarrow |R_{v}\rangle$ 
respectively with $p=\pm1$ exhibiting the maximal interference among the decay paths. Here $p=1$ corresponds to the two dipole moment vectors parallel to each other on the other hand when they are anti-parallel $p=-1$. Non-orthogonal dipole moments in optical transition have been generated using superposition of singlet and triplet states due to spin-orbit coupling in sodium dimers~\cite{Xia97}. More generally, interference strength $p$ is a phase shift acquired by wavefunction between initial and final states. Equations (\ref{eq:a2})-(\ref{eq:b}) 
can be written and solved in the dressed basis using the approach developed by Scully \cite{Scully} as discussed in Appendix A for general $p$ and in the presence of additional decay rates $\Gamma, \gamma$. The probability of emission $P_{emiss}$ 
defined as a sum of population of the doublet $|v_1\rangle$, $|v_2\rangle$ and of the reservoir state $|R_v\rangle$ 
due to conservation of probability,  can be written in terms of populations of states $|c\rangle$ and $|R_c\rangle$ as
\begin{equation}\label{eq:emK}
\begin{split}
P_{\text{emiss}}(\tau|c)=1-|c(\tau)|^{2}-2\tilde{\Gamma}\int_0^\tau|c(\tau')|^{2}d\tau'.
\end{split}
\end{equation}
In the long time limit, $\tau\gg1,1/\tilde{\Gamma}$ and assuming $\gamma_1=\gamma_2=\gamma$ for simplicity, the 
probability of emission defined in Eq.(\ref{eq:emK}) (derived in Appendix B) yields
 \begin{equation}\label{eq:Pemiss0}
P_{\text{emiss}}(\infty|c)=\frac{(\tilde{\Gamma}+1)(\tilde{\Omega}_1^2+\tilde{\Omega}_2^2)-2p\tilde{\Omega}_1\tilde{\Omega}_2}{\tilde{\Gamma}\left[(\tilde{\Gamma}+1)^{2}+\tilde{\Delta}^2-p^{2}\right]}.
\end{equation}
where the tilde signifies that all the parameters are now dimensionless as they are normalized by $\gamma$. 
The probability of absorption from level $|v_1\rangle$ can be evaluated in a similar manner. For the 
initial conditions $v_1(0)=1$,$v_2(0)=0$ and $c(0)=0$ the probability of absorption $P_{abs}$ is given by the sum of
population on states $|c\rangle$ and $|R_c\rangle$:
  \begin{equation}\label{eq:prho1}
 P_{\text{abs}}(\tau|v_1)=|c(\tau)|^{2}+2\tilde{\Gamma}\int_{0}^{\tau}|c(\tau')|^{2}d\tau'
 \end{equation}
that yields the following expression in the long time limit, $\tau\gg1,1/\tilde{\Gamma}$ (see Appendix B)

 \begin{equation}\label{eq:Pabs01}
 \begin{split}
P_{\text{abs}}(\infty|v_1)=\frac{1}{\mathcal{D}}\left\{\left[2(1+\tilde{\Delta}^{2})(1+\tilde{\Gamma})-\tilde{\Gamma} p^{2}\right]\tilde{\Omega}_1^2 \right. \\
\left. -2(\tilde{\Gamma}+2)p\tilde{\Omega}_1\tilde{\Omega}_2+(\tilde{\Gamma}+2)p^{2}\tilde{\Omega}_2^2\right\}
\end{split}
\end{equation}
where $\mathcal{D}= 2(1+\tilde{\Delta}^2-p^{2})\left[(\tilde{\Gamma}+1)^{2}+\tilde{\Delta}^2-p^{2}\right]$. 
The probability of absorption from level $|v_2\rangle$ can be derived in the same way as for the level $|v_1\rangle$ 
by interchanging
$v_1\leftrightarrow v_2$ in Eq. (\ref{eq:prho1}) and $\tilde{\Omega}_1\leftrightarrow \tilde{\Omega}_2$ in 
Eq. (\ref{eq:Pabs01}). Comparison of  Eq. (\ref{eq:Pemiss0}) with Eq. (\ref{eq:Pabs01}) yields that probability 
of emission and absorption can vary substantially in the presence ($p\neq 0$) or absence ($p=0$) of interference. 

So far we have discussed a model with doublet in the ground state.  Let us now consider doublet in the excited 
state (as shown in Fig. 1(b)). In practice this configuration is commonly used in semiconductor systems like quantum wells
and dots. The expression for the probability of emission and absorption in case of excited state doublet can 
be obtained as follows. If we start with $|c_{1}\rangle$, the probability of emission is given by 
  \begin{equation}\label{Uemiss}
 P_{\text{emiss}}(\tau|c_1)=|v(\tau)|^{2}+2\tilde{\Gamma}\int_{0}^{\tau}|v(\tau')|^{2}d\tau'
 \end{equation}
Similarly the probability of absorption from $|v\rangle$ yields 
\begin{equation}\label{Uabs}
\begin{split}
P_{\text{abs}}(\tau|v)=1-|v(\tau)|^{2}-2\tilde{\Gamma}\int_0^\tau|v(\tau')|^{2}d\tau'.
\end{split}
\end{equation}
The expression for the emission and absorption probability can be calculated by following 
a procedure similar to that outlined in appendix B for the ground state doublet. In the long time limit 
$t \gg \gamma^{-1}, \Gamma^{-1}$, we find that the expression for 
emission and absorption probabilities obtained from Eqs.(\ref{Uemiss})-(\ref{Uabs}) reduces 
to Eqs.(\ref{eq:Pabs01}) and (\ref{eq:Pemiss0}) respectively.
\section{Discussion}
\textit{Applications to Lasing without Inversion and Photodetectors}: The model discussed in the previous section 
is relevant for the design of the systems with nonrecpirocal relation between emission and absorption. For instance, 
suppressed absorption or/and enhanced emission in the laser systems allows for operating without population inversion. 
On the other hand enhanced absorption with suppressed emission can results in the photodetector or photovoltaic/solar cell system with 
enhanced power output \cite{Dor11, Dor111}. Both LWI and photodetector schemes can be realized in atomic 
molecular and semiconductor systems. In atoms Agarwal-Fano type QI can arise between decay channels from magnetic sub-levels. 
In molecular systems on the other hand, decay pathways of different vibrational/rotational levels lead to asymmetric 
absorption/emission profiles due to interference. In the case of semiconductors, Agarwal-Fano interference 
comes about quite naturally in a system of  two quantum wells or dots grown at nanometer separations \cite{Cap97, Sch97}.
The tunneling/F\"{o}rster interactions among
the wells/dots renormalizes the bare energies and bare states of the system thereby creating new eigenstates which then 
reveals the interference in decay channels through tunneling to the same continuum \cite{Sit11, Scully}.
Note that QI and coherence effects in semiconductors are strongly effected by the presence of
dephasing environment and hence experiments in these systems are carried out at very low temperatures (10 K). 
This thereby restrict their practical feasibility for various applications involving QI. 
However, recently quantum dot photodetector enhanced by Fano-type interference assisted with metallic hole 
array was reported operating at 77 K \cite{Vas11}. Hence in near future realization of Fano like QI effects in 
nanostructure and its various applications might be achievable even at room temperatures.

To put the above ideas to prospective, we discuss the functional dependence of the of emission and absorption 
probabilities on the interference strength $p$ and the level spacing $\Delta$ in the steady state and 
transient regime. We show in Fig. 2 the steady state behavior and temporal evolution of emission and absorption probabilities for different values of $p$ and  $\Delta$.
Figures in the upper panel (2a, b, c) correspond to large level spacing compare 
to spontaneous decay rate $\Delta\gg\gamma$ ($\tilde{\Delta}\gg 1$). The steady state emission profile is seen to be strongly influenced by the strength of QI. It varies from its minimum at $p=1$ to maximum at $p=-1$ (see Fig. 2a). The enhancement in emission is found to be almost 10 fold. However for absorption the effect of interference is not significant as $p$ varies from $-1$ to $1$. Therefore, for $p= -1$ one can achieve regime with largest emission, which can be useful in inversionless lasing schemes. On the other hand at $p=1$, as emission reaches its minimum, it is attractive in realization of photo-detectors and photovoltaic devices. Note, that in semiconductor double quantum well system, control over $p$ can be achieved by manipulating of the width of the shallow well \cite{Sch97}. The time evolution of the resonance profiles shown in Fig. 2b and 2c exhibits oscillatory behavior in the emission and absorption probabilities. Period of oscillations is determined by the frequency $\sqrt{\Delta^2-\gamma^2}$ and thus strongly depend on the level spacing. We see further that the oscillations gets damped with time and the probabilities eventually reaches the steady state. 

For small level spacing $\Delta\ll\gamma$ ($\tilde{\Delta}\ll 1$), the situation becomes less trivial. In this case the behavior of emission and absorption profiles is depicted in the lower panel of Fig. 2 (d, e, f)). In the steady state the both the probabilities varies significantly with the interference strength $p$ (see Fig. 2d). We find that while absorption probability increases monotonically from $p = 1$ to $p = -1$, emission is seen to first increase until about $p = -0.5$ beyond which it rapidly decreases to reach the minimum value at $p = -1$. This is in sharp contrast to the behavior of the emission probability for large $\Delta$. In the time dependent profiles (Fig. 2e, 2f) we find that in comparison to the case of large splitting both emission and absorption probabilities show no oscillations and  reach their steady state values that depend strongly on the interference strength. Furthermore, interesting case arise at $p=-1$ where emission profile first reaches its maximum and then drops down to the steady state that has the smallest value compare to other $p\neq -1$. In the same time absorption profile at $p=-1$ reaches its maximum value at steady state. Note that in contrast to that, for large splitting at $p=-1$ emission has its maximum (see Fig. 2b). Therefore, not only interference strength determines the emission and absorption profile, but the level spacing itself has strong impact. Namely, for fixed value of $p$, for example $p=-1$, large level spacing $\Delta$ yields  the strongest emission (see Fig. 2b) which is in favor of lasing process. In the same time for small level spacing the emission is strongly suppressed while absorption reaches its maximum (See Figs. 2e,f), which is perfect situation for photo-detection and photocell operation. Furthermore, it is worth noting, that despite the asymmetry between curves for $p=\pm 1$ in Fig. 2, result for $p=1$ can be derived from $p=-1$ case by changing the sign 
of the Rabi frequency, for instance: $\Omega_1\to-\Omega_1$. 

\begin{figure}[t]\label{fig:ratio}
\centerline{\includegraphics[height=4.5cm,width=0.46\textwidth,angle=0]{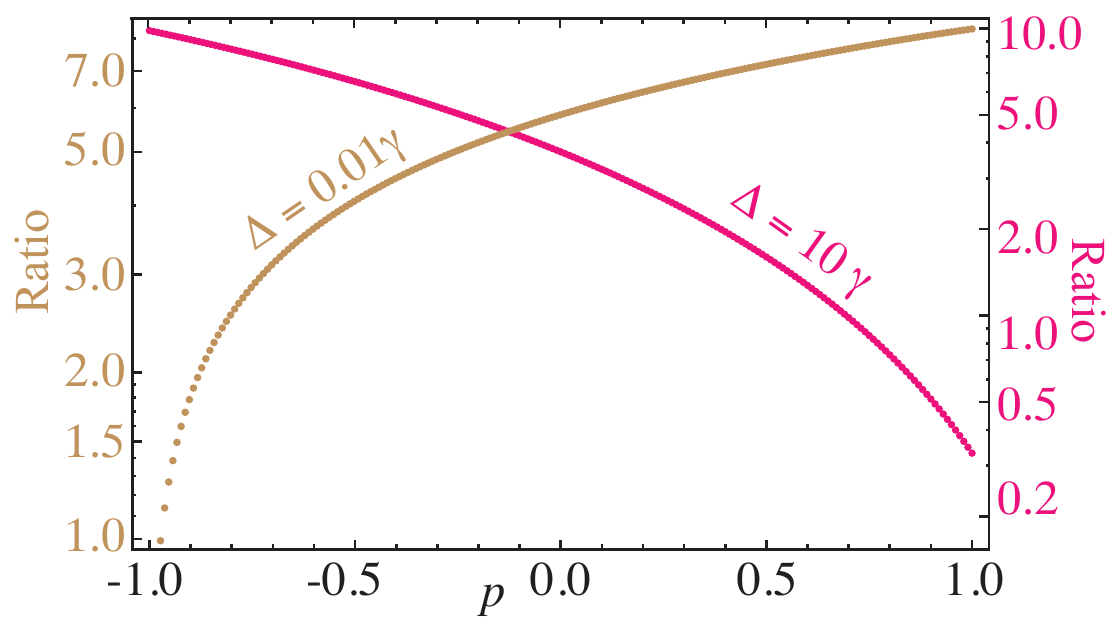}}
\caption{(Color online) Ratio of the probability of emission to absorption for two combinations of coupling $\Delta$ as a function of 
the parameter $p$. For numerical simulation we took, $\Omega_{1}=\Omega_{2}=0.3\gamma$, $\Gamma=0.4\gamma$.}
\end{figure}

To study further the effects of $p$ and $\Delta$ and to understand the special case of antiparallel 
alignment $p=-1$ consider the ratio of emission and absorption given by Eq. (\ref{eq:Pemiss0}) and (\ref{eq:Pabs01}):
\begin{equation}\label{eq:rat}
\frac{P_{emiss}}{P_{abs}}=\frac{2(1+\tilde{\Gamma}-p)(1+\tilde{\Delta}^2-p^2)}{\tilde{\Gamma}
[\tilde{\Delta}^2(1+\tilde{\Gamma})+\tilde{\Gamma}(1-p)+(1-p)^2]},
\end{equation}
where for simplicity we assume $\Omega_1=\Omega_2$. Fig. 3 shows the ratio in Eq. (\ref{eq:rat}) as a 
function of interference strength $p$ for the case of small and large level spacing. If the spacing is small,
$\Delta\ll\gamma$, then the ratio in Eq. (\ref{eq:rat}) monotonically increasing from $p=-1$ to $p=1$, 
while for large spacing $\Delta\gg\gamma$, the behavior is essentially the opposite, i.e. it is 
monotonically decreasing function as we mention above. Furthermore, in the limit of weak field 
$\Omega_1=\Omega_2=\Omega\ll 1$ Eq. (\ref{eq:rat}) yields for $p=0,1$ result that is independent of $\Delta$. 
Namely for no interference, i.e. $p=0$ Eq. (\ref{eq:rat}) yields $2/\tilde{\Gamma}$, while 
for parallel alignment $p=1$ it yields $2/(1+\tilde{\Gamma})$. On the other hand the case of 
antiparallel alignment ($p=-1$) is special. In particular, for small spacing $\Delta\ll\Gamma\ll\gamma$ 
Eq. (\ref{eq:rat}) gives $\tilde{\Delta}^2/\tilde{\Gamma}\ll 1$, while for $\Delta\gg\gamma$ 
and $\Gamma\ll\gamma$ the result is $4/\tilde{\Gamma}\gg1$. Therefore, the present 
analysis not only confirms that destructive interference can alter the detailed balance but also 
exhibits that by controlling two  parameters. Namely  by adjusting the interference strength $p$ and energy spacing $\Delta$, one can regulate
the ratio between emission and absorption probabilities in the system. This possible manipulation of 
$p$ and $\Delta$ hence also suggest that in the same system with two lower (upper) levels one can induce either suppression of  emission \cite{Dor11, Dor111} or absorption \cite{Cap97, Sch97}, respectively. The later choice governed by level spacing $\Delta$ can be also controlled externally either by adjusting the current through the junction, or by manipulating the magnetic field in hyperfine splitting~\cite{Jha1,Jha2}. In Fig 4. we have plotted the effect of $\Delta$ on the temporal evolution of the probability of emission. The results show that the oscillations in the probability varies with the increase of $\Delta$. Furthermore, for fixed $\Delta$ and $\gamma$ the number of oscillations is governed by rate $\Gamma$ since probability decays as $\exp(-\Gamma t)$. For interference strength $p$, control can be achieved by a tailored variation of the quantum well widths \cite{Sch97}. Summarizing the proposed scheme with lower doublet can be applied to the system that requires emission (absorption) suppression or enhancement and thus  is very attractive for both: light emitting devices, such as LWI  and light absorbing photodetector 
systems. 
\begin{figure}[t]
\centerline{\includegraphics[height=4.5cm,width=0.46\textwidth,angle=0]{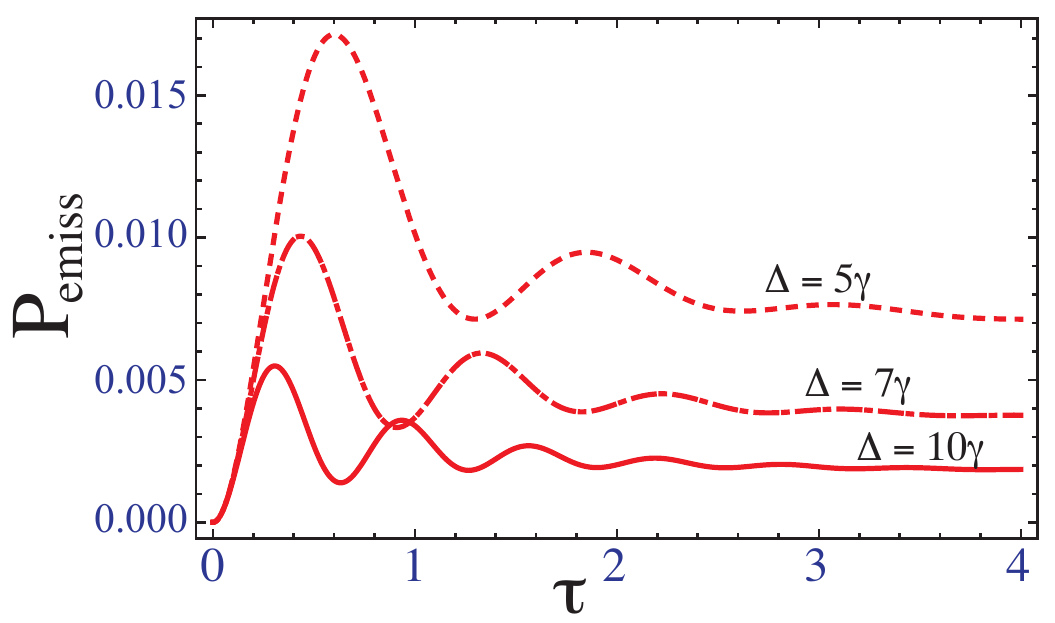}}
\caption{(Color online) Probability of emission in the three level model lower doublet for different choices of $\Delta$. 
For numerical simulation we took, $\Omega_{1}=\Omega_{2}=0.3\gamma$, $\Gamma=0.4\gamma, \gamma=1, \tau= \gamma t, p=1$.}
\end{figure}

\textit{Quantum beats in semiclassical picture :} Besides broad range of applications, interference effects 
and in particular its sensitivity to the level spacing  discussed in the present work are related to 
fundamental question about the  applicability of semiclassical theory in quantum problems. Semiclassical 
description (SCT) can predict  self-consistent and physically acceptable behavior of many physical systems 
and explain almost all quantum phenomena. However It is not always correct. For instance, the phenomena 
of quantum beats has substantially different result if considered in the framework of quantum 
electrodynamics (QED)\cite{Scub}. Namely, for different configurations of three-level systems: 
for instance $V$and $\Lambda$ schemes (see Fig.  5) that are initially prepared in a coherent superposition 
of all three states SCT description predicts the existence of quantum beats for both schemes, 
whereas QED theory predicts no quantum beats in the case of $\Lambda$ scheme.  The explanation 
of the phenomenon is quite straightforward and based on quantum theory of measurements. 
In the case of $V$ scheme the coherently excited atom decays to the same final state $|v\rangle$ starting 
from $|c_+\rangle$ and $|c_-\rangle$ and one cannot determine which decay channel was used. 
Therefore this interference that is similar to the double-slit problem leads to the existence of 
quantum beats. However in the case of $\Lambda$ scheme that has also two decay 
channels: $|c\rangle \to|v_+\rangle$ and $|c\rangle\to|v_-\rangle$, after a long 
time the observation of the atom's final state ($|v_+\rangle$ or $v_-\rangle$)
will determine which decay channel was used. In this case we do not expect quantum beats. 
Three-level systems with doublet in the ground state or excited state is in a way similar 
to the $\Lambda$ and $V$ types of atom respectively. Therefore we can also study the 
quantum beats effect in those systems. Note that in the model of Fig. 1 we have 
additional radiative decays of states which guarantees that system can reach a 
steady state within finite amount of time.
\begin{figure}[t]\label{fig:qbeats}
\centerline{\includegraphics[height=4.5cm,width=0.4\textwidth,angle=0]{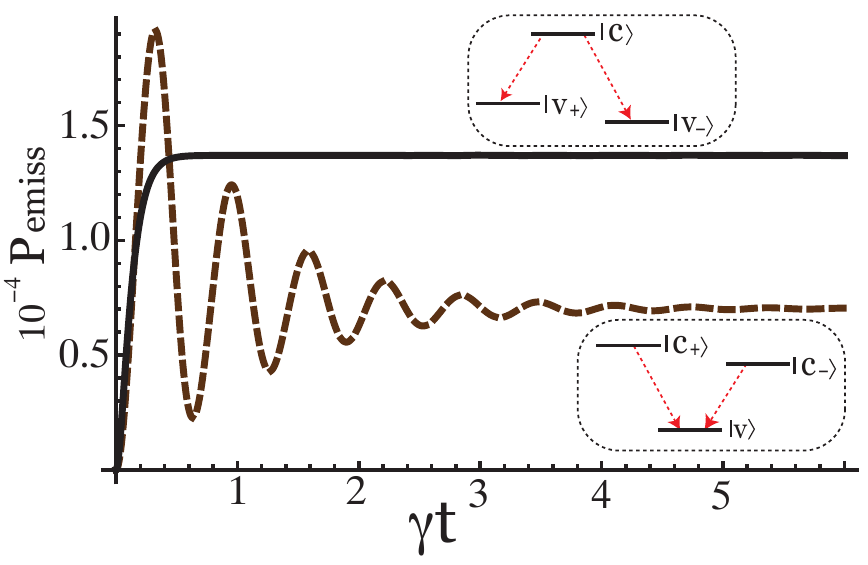}}
\caption{(Color online) Probability of emission $P_{\text{emis}}$ as a function of dimensionless time $\tau$ 
for three-level system with doublet in excited state - (dashed line) and for three level system 
with doublet in ground state - (solid line) calculated numerically according to Eqs. (\ref{eq:emK}) 
and (\ref{eq:prho1}) based on the solution of Eqs. (\ref{eq:a2})- (\ref{eq:b}). For numerical 
simulations we took $\Omega_{1}=0.1\gamma, \Omega_{2}=0.08\gamma, \Gamma=10\gamma, \Delta=0.1\gamma$.}
\end{figure}
Fig. 5 illustrates that in the case of doublet in excited state ($V$ scheme) with large spacing 
between levels $|c_+\rangle$ and $|c_-\rangle$ $\tilde{\Delta}>>1$, the probability of emission 
oscillates as  a function of time and reaches the steady state at the time scale determined by 
radiative decay $1/\tilde{\Gamma}>>1$. However, for the case of doublet in the ground state ($\Lambda$ scheme) 
with small spacing $\tilde{\Delta}<<1$ the probability of emission does not process any quantum beats 
and smoothly reach the steady state. Therefore, phenomenon of Fano interference has a potential to 
resolve the fundamental question  about an applicability of the  semiclassical description to the 
problem of quantum beats.

\section{Conclusion}
To conclude, in this paper we investigated the effect vacuum induced QI on the emission(absorption) profile of a three-level system with a doublet in the ground  or excited state (see Fig. 1(a)). We show that QI can enhance the balance breaking between emission and absorption. We use probability amplitude method, since the states involved in calculation have zero photon occupation number. Furthermore, our findings are in full agreement with the results obtained by density matrix formalism. We observed that the interference strength $p$ governed by the phase shift between the decay pathways play a crucial role on the emission(absorption) dynamics of the system. For the closely spaced doublet $(\Delta \ll \gamma)$, for which the vacuum induced QI becomes important, the behavior of the emission(absorption) profile of our model appears counterintuitive. For $p \sim -1$, the ratio of probability of emission to probability of absorption is very small, a condition favorable for applications like photovoltaics. On the other hand for $p \sim 1$, the ratio is large thus favorable for amplification without population inversion in steady-state (see Fig 2 (b,e)). In addition to these applications we found that Agarwal-Fano QI can also predicts the occurrence of fundamental phenomena like quantum beats in the semi-classical framework, that fully agrees with the QED description.  

\section{Acknowledgment}
We thank Anatoly Svidzinsky and Dong Sun  for useful and stimulating discussion and acknowledge the support from the Office of Naval Research, Robert A. Welch Foundation (Award A-1261). P.K.J also acknowledges Herman F. Heep and Minnie Belle Heep Texas A$\&$M University Endowed Fund held/administered by the Texas A$\&$M Foundation.

\appendix
\section{The Scully dressed state analysis}\label{app:diag}

We start with evolution of amplitudes in Eqs.~(\ref{eq:a2})-(\ref{eq:b}) for $\gamma_1=\gamma_2=\gamma$

\begin{equation}
\dot{v}_2=-(\gamma+i\Delta)v_2-p\gamma v_1-i\Omega_2c \label{a2},
\end{equation}

\begin{equation}
\dot{v}_1=-(\gamma-i\Delta)v_1-p\gamma v_2-i\Omega_1c \label{a1},
\end{equation}

\begin{equation}
\dot{c}=-i\Omega_2v_2-i\Omega_1v_1 \label{b}-\Gamma c,
\end{equation}
Writing Eqs.~(\ref{a2})-(\ref{b}) in matrix form, we obtain
\begin{equation}
\frac{d}{d\tau} \left(
  \begin{array}{c}
    v_2 \\
    v_1 \\
    c \\
  \end{array}
\right) =-\Gamma_0 \left(
                   \begin{array}{c}
                     v_2 \\
                     v_1 \\
                     c \\
                   \end{array}
                 \right)-iV\left(
                             \begin{array}{c}
                               v_2 \\
                               v_1 \\
                               c \\
                             \end{array}
                           \right),
\end{equation}

where $\tau = \gamma t $, and the Fano decay matrix is defined by
\begin{equation}
\Gamma_0=
\left(
  \begin{array}{ccc}
    1+i \tilde{\Delta} & p & 0 \\
    p & 1-i \tilde{\Delta} & 0 \\
    0 & 0 & \Gamma \\
  \end{array}
\right),
\end{equation}

and probe-field interaction is given by

\begin{equation}
V= \left(
  \begin{array}{ccc}
    0 & 0 & \tilde{\Omega}_2 \\
    0 & 0 & \tilde{\Omega}_1 \\
    \tilde{\Omega}_2 & \tilde{\Omega}_1 & 0 \\
  \end{array}
\right),
\end{equation}

with $\tilde{\Delta}=\frac{\Delta}{\gamma}$ and $
\tilde{\Omega}_{1,2}= \frac{\Omega_{1,2}}{\gamma}$.

It is intuitive to introduce a basis in which the Fano coupling is
transformed away. We proceed from the bare basis via the $U$,
$U^{-1}$ matrices of diagonalization.

\begin{equation}
U^{-1}=\frac{1}{\sqrt{2}p} \left(
                           \begin{array}{ccc}
                             p & p & 0 \\
                             x-i \tilde{\Delta} & -x-i \tilde{\Delta} & 0 \\
                             0 & 0 & \sqrt{2}p \\
                           \end{array}
                         \right),
\end{equation}
\begin{equation}
U=\frac{1}{\sqrt{2}x} \left(
                           \begin{array}{ccc}
                             x+ i \tilde{\Delta} & p & 0 \\
                             x-i \tilde{\Delta} & -p & 0 \\
                             0 & 0 & \sqrt{2}x \\
                           \end{array}
                         \right).
\end{equation}
Here $x=\sqrt{p^2-\tilde{\Delta}^2}$.

so that the transformed state vector is defined by
\begin{equation}
U\left(
   \begin{array}{c}
     v_2 \\
     v_1 \\
     c \\
   \end{array}
 \right) =\left(
            \begin{array}{c}
              V_+ \\
              V_- \\
              C \\
            \end{array}
          \right),
\end{equation}

which implies

\begin{equation}\label{eq:tran}
V_{\pm}=\frac{(x\pm i\tilde{\Delta})v_2\pm pv_1}{\sqrt{2}x}
\end{equation}

and thus,

\begin{equation}
\left(
            \begin{array}{c}
              \dot{V}_+ \\
              \dot{V}_- \\
              \dot{C} \\
            \end{array}
   \right)
   =-\Gamma_t \left(
            \begin{array}{c}
              V_+ \\
              V_- \\
              C \\
            \end{array}
          \right)-iV_t \left(
            \begin{array}{c}
              V_+ \\
              V_- \\
              C \\
            \end{array}
          \right),
\end{equation}

in which the diagonal $\Gamma_t$ operator is
\begin{equation}
\Gamma_t=U\Gamma_0 U^{-1}=\left(
                          \begin{array}{ccc}
                            1+x & 0 & 0 \\
                            0 & 1-x & 0 \\
                            0 & 0 & 
                            \Gamma \\
                          \end{array}
                        \right),
\end{equation}

and the transformed interaction potential is
\begin{widetext}
\begin{equation}
V_t= UVU^{-1}=\frac{1}{\sqrt{2}p}\left(
  \begin{array}{ccc}
    0 & 0 & p[\tilde{\Omega}_2(x+i \tilde{\Delta})+p\tilde{\Omega}_1]/x \\
    0 & 0 & p[\tilde{\Omega}_2(x-i \tilde{\Delta})-p\tilde{\Omega}_1]/x \\
    \tilde{\Omega}_2+\tilde{\Omega}_1 (x-i\tilde{\Delta}) & \tilde{\Omega}_2-\tilde{\Omega}_1 (x+i\tilde{\Delta}) & 0 \\
  \end{array}
\right).
\end{equation}
\end{widetext}

The equation of motion in terms of $V_\pm$ and $C$ are then found to
be

\begin{equation}
\frac{d
V_+}{d\tau}=-(1+x)V_+-\frac{i}{\sqrt{2}x}[\tilde{\Omega}_2(x+i
\tilde{\Delta})+p\tilde{\Omega}_1]C, \label{dAp}
\end{equation}

\begin{equation}
\frac{d
V_-}{d\tau}=-(1-x)V_--\frac{i}{\sqrt{2}x}[\tilde{\Omega}_2(x-i
\tilde{\Delta})-p\tilde{\Omega}_1]C, \label{dAn}
\end{equation}

\[
\frac{d
C}{d\tau}=-\tilde{\Gamma}C-\frac{i}{\sqrt{2}}[p\tilde{\Omega}_2+\tilde{\Omega}_1(x-i\tilde{\Delta})]V_+-
\]
\begin{equation}
-\frac{i}{\sqrt{2}}[p\tilde{\Omega}_2-\tilde{\Omega}_1(x+i\tilde{\Delta})]V_-,
\label{adB}
\end{equation}

\section{Derivation of emission and absorption probabilities in dressed basis }\label{app:emabs}

We start with amplitude equations in dressed basis (\ref{dAp}) - (\ref{adB}). The initial conditions corresponding to 
the emission from the state $C$ are $V_{\pm}(0)=0$, $C(0)=1$. Assuming the driving fields to be weak 
($\tilde{\Omega}_{1,2}<<1$ we can solve Eqs. (\ref{dAp}) - (\ref{adB}) by expansion in perturbation series 
over $\tilde{\Omega}_{1,2}$. The lowest order solution for $B(\tau)$ of Eq. (\ref{adB}) yields 
$C^{(0)}(\tau)=e^{-\tilde{\Gamma}\tau}$. The latter can be substituted in Eqs. (\ref{dAp}) and (\ref{dAn}) 
to find $V_{\pm}^{(0)}(\tau)$:

\begin{equation}\label{eq:apm}
V_{\pm}^{(0)}(\tau)=-i\frac{\tilde{\Omega}_2(x\pm i\tilde{\Delta})\pm p\tilde{\Omega}_1}{\sqrt{2}x(1\pm x-\tilde{\Gamma})}\left(e^{-\tilde{\Gamma}\tau}-e^{-(1\pm x)\tau}\right)
\end{equation}
The exponential approximation or $C(\tau)$ gives relatively good agreement with numerical simulations only for small time. 
For large time the behavior of the system is far from being exponential. Therefore, we should consider next order 
correction for $C(\tau)$. It can be done by substituting  functions $V^{(0)}_{\pm}$ from Eq. (\ref{eq:apm}) to Eq. (\ref{adB}) 
which yields 
\[
C^{(1)}(\tau)=\left[\frac{A_+}{1+x-\tilde{\Gamma}}+\frac{A_-}{1-x-\tilde{\Gamma}}-(A_++A_-)\tau\right]e^{-\tilde{\Gamma}\tau}
\]
\begin{equation}\label{eq:B1}
+e^{-\tilde{\Gamma}\tau}-\frac{A_+}{1+x-\tilde{\Gamma}}e^{-(1+x)\tau}-\frac{A_-}{1-x-\tilde{\Gamma}}e^{-(1-x)\tau},
   \end{equation}
   where
   \begin{equation}
   A_{\pm}=\frac{[p\tilde{\Omega}_2\pm(x\mp i\tilde{\Delta})\tilde{\Omega}_1][\tilde{\Omega}_2(x\pm i\tilde{\Delta})\pm p\tilde{\Omega}_1]}{2px(1\pm x-\tilde{\Gamma})}.
   \end{equation}
  Using the definition for emission probability from Eq. (\ref{eq:emK}) at large time $\tau\gg 1,1/\tilde{\Gamma}$, 
neglecting higher order terms in $\tilde{\Omega}_{1,2}$  the probability of absorption yields \begin{equation}\label{eq:Pemis}
P_{\text{emiss}}(\infty|b)=\frac{(\tilde{\Gamma}+1)(|\tilde{\Omega}_1|^2+|\tilde{\Omega}_2|^2)-2p\tilde{\Omega}_1
\tilde{\Omega}_2}{\tilde{\Gamma}\left[\tilde{\Delta}^2+(\tilde{\Gamma}+1)^2-p^2\right]}.
\end{equation} 

Similarly one can derive the probability of absorption. We start from absorption from level $v_1$. The initial 
conditions for system with population on $v_1$ in dressed states are $V_{\pm}(0)=\pm p/\sqrt{2}x$, $C(0)=0$ 
(see Eq. (\ref{eq:tran})). In lowest order of $\tilde{\Omega}_{1,2}$, Eqs. (\ref{dAp}) and (\ref{dAn}) yield
 \begin{equation}\label{eq:apm0em}
V_{\pm}^{(0)}(\tau|v_1)=\pm\frac{p}{\sqrt{2}x}e^{-(1\pm x)\tau}.
\end{equation} 
Corresponding zero order solution of $C^{(0)}(\tau)$ of Eq. (\ref{adB}) is given by 
 \begin{equation}\label{eq:bem}
C^{(0)}(\tau|v_1)=B_+e^{-(1+x)\tau}-B_-e^{-(1-x)\tau}+(B_--B_+)e^{-\tilde{\Gamma}\tau},
\end{equation}
where
\begin{equation}
B_{\pm}=i\frac{p\tilde{\Omega}_2\pm\tilde{\Omega}_1(x\mp i\tilde{\Delta})}{2x(1\pm x-\tilde{\Gamma})}
\end{equation}
Therefore, probability of absorption form level $v_1$ for large time $\tau\gg 1,1/\tilde{\Gamma}$ given by 
Eq. (\ref{eq:prho1}) reads
\[
P_{\text{abs}}(\infty|v_1)=
\]
 \begin{equation}\label{eq:Pabs1}
\frac{(\tilde{\Gamma}+2)|\tilde{\Omega}_1-p\tilde{\Omega}_2|^2+[\tilde{\Gamma}(1-p^2)+2\tilde{\Delta}^2(\tilde{\Gamma}+1)]\tilde{|\Omega_1|^2}}{2(1+\tilde{\Delta}^2-p^2)[\tilde{\Delta}^2+(\tilde{\Gamma}+1)^2-p^2]}
\end{equation}

The probability of absorption from level $v_2$ can be derived in the same way as for the level $v_1$. In 
this case, the initial conditions according to Eq. (\ref{eq:tran}) read $V_{\pm}(0)=(x \pm i\tilde{\Delta})/
\sqrt{2}x$, $C(0)=0$. In lowest order of $\tilde{\Omega}_{1,2}$, Eqs. (\ref{dAp}) and (\ref{dAn}) have the 
following solution:
 \begin{equation}\label{eq:apm0em2}
V_{\pm}^{(0)}(\tau|a_2)=\frac{x\pm i\tilde{\Delta}}{\sqrt{2}x}e^{-(1\pm x)\tau}.
\end{equation} 
Corresponding zero order solution of $C^{(0)}(\tau)$ of Eq. (\ref{adB}) yields
\begin{equation}
C^{(0)}(\tau|v_1)=D_+e^{-(1+x)\tau}+D_-e^{-(1-x)\tau}+(D_++D_-)e^{-\tilde{\Gamma}\tau},
\end{equation}
where
\begin{equation}
D_{\pm}=i\frac{[p\tilde{\Omega}_2\pm\tilde{\Omega}_1(x\mp i\tilde{\Delta})](x\pm i\tilde{\Delta})}{2px(1\pm x-\tilde{\Gamma})}
\end{equation}

Therefore, probability of absorption from level $v_2$  for $\tau\gg 1,1/\tilde{\Gamma}$ given by Eq. (\ref{eq:prho1}) yields
 \[
P_{\text{abs}}(\infty|v_2)=
\]
 \begin{equation}\label{eq:Pabs2}
\frac{(\tilde{\Gamma}+2)|\tilde{\Omega}_2-p\tilde{\Omega}_1|^2+[\tilde{\Gamma}(1-p^2)+2\tilde{\Delta}^2
(\tilde{\Gamma}+1)]\tilde{|\Omega_2|^2}}{2(1+\tilde{\Delta}^2-p^2)[\tilde{\Delta}^2+(\tilde{\Gamma}+1)^2-p^2]},
\end{equation}
which becomes Eq. (\ref{eq:Pabs1}) if $\tilde{\Omega}_1\leftrightarrow\tilde{\Omega}_2$.

\end{document}